\documentclass[reprint,eqsecnum,floats,aps,amsmath,amssymb,nofootinbib,prd,onecolumn,showpacs]{revtex4-2}
\usepackage{bm}
\usepackage{graphicx,physics}
\usepackage{amsmath,amssymb,mathtools,mathrsfs}
\usepackage{hyperref}
\usepackage{graphicx}
\usepackage{arydshln}
\usepackage{xcolor}
\usepackage{braket}
\usepackage{tensor}
\usepackage{enumitem,array,textcomp}

\begin{document}

\title{Extended phase space quantization of a black hole interior model in Loop Quantum Cosmology}

\author{Beatriz Elizaga Navascu\'es}
\email{Electronic addres: bnavascues@lsu.edu}
\affiliation{Department of Physics and Astronomy, Louisiana State University, Baton Rouge, LA 70803-4001, USA}
\author{Guillermo A. Mena Marug\'an}
\email{Electronic addres: mena@iem.cfmac.csic.es}
\affiliation{Instituto de Estructura de la Materia, IEM-CSIC, C/ Serrano 121, 28006 Madrid, Spain}
\author{Andr\'es M\'{\i}nguez-S\'anchez}
\email{Electronic addres: andres.minguez@iem.cfmac.csic.es}
\affiliation{Instituto de Estructura de la Materia, IEM-CSIC, C/ Serrano 121, 28006 Madrid, Spain}
\begin{abstract}

Considerable attention has been paid to the study of the quantum geometry of nonrotating black holes within the framework of Loop Quantum Cosmology. This interest has been reinvigorated since the introduction of a novel effective model by Ashtekar, Olmedo, and Singh. Despite recent advances in its foundation, there are certain questions about its quantization that still remain open. Here we complete this quantization taking as starting point an extended phase space formalism suggested by several authors, including the proposers of the model. Adopting a prescription that has proven successful in Loop Quantum Cosmology, we construct an operator representation of the Hamiltonian constraint. By searching for solutions to this constraint operator in a sufficiently large set of dual states, we show that it can be solved for a continuous range of the black hole mass. This fact seems in favour of a conventional classical limit (at least for large masses) and contrasts with recent works that advocate a discrete spectrum. We present an algorithm that determines the solutions in closed form. To build the corresponding physical Hilbert space and conclude the quantization, we carry out an asymptotic analysis of those solutions, which allows us to introduce a suitable inner product on them.

\end{abstract}

\maketitle

\section{Introduction\label{sec: I}}

General Relativity (GR) and Quantum Mechanics play a central role in modern physics. However, these theories are based on different physical and logical principles. To overcome this tension, various proposals have been made to construct a theory of quantum gravity (see e.g. Ref. \cite{QG}). Among them, Loop Quantum Gravity (LQG) is one of the most solid candidates. The foundations of LQG rest upon a canonical, nonperturbative quantization of GR that is independent of background structures. This canonical formulation is achieved by means of a 3+1 decomposition of the spacetime. In this framework, the holonomies of the Ashtekar-Barbero SU(2) connection along closed loops and the fluxes of densitized triads through surfaces serve as the fundamental (gauge-invariant) functions on phase space, facilitating the quantization process \cite{A&L, Thiemann}. The selection of the quantum representation for these variables is a pivotal aspect in the construction of the theory and stands out as one of the distinctive features of LQG, notably diverging from the representations used in other quantum field theories, such as the standard Fock representations \cite{A&L, Thiemann}. Although there have been promising advances, the quantization program posed by LQG has not yet been completed owing to the highly intricate nature of the Hamiltonian constraint, that encapsulates the time reparametrization invariance of GR. Nevertheless, to progress towards this completion, LQG techniques have been extensively applied to cosmological studies \cite{A&S, GMM, APS, FMO}, providing a suitable arena to check the quantization methods and extract predictions. This has given rise to the discipline known as Loop Quantum Cosmology (LQC).

High curvature scenarios, such as the Early Universe or black holes, are in fact crucial for testing quantum gravity, as its effects are expected to become significant in these situations. In particular, LQC has successfully been used to investigate a variety of cosmological models, both isotropic and anisotropic, with and without matter content, and with and without spatial curvature \cite{A&S,GMM}. It has also led to predictions about the effects of LQG on the primordial spectra of cosmological perturbations (see e.g. Refs. \cite{AAN1,AAN2,Ivan,hyb1,hyb2,hyb-review,hyb-others,AshNe}). In contrast, the application of LQG to the study of black hole spacetimes is not so firmly established yet, in spite of the considerable attention paid to this problem (for a far from exhaustive list of works on this issue, see Refs.\cite{LBH1,LBH2,A&B,B&V,Chiou,LBH3,LBH4,LBH5,JPS,LBH6,LBH7,LBH8,LBH9,LBH10,LBH11,LBH12}). 

The motivation for investigating the quantum behavior of black holes is manifold. It is expected to be important in order to elucidate the nature of Hawking radiation in dynamical black holes beyond semiclassical approximations, and eventually to shed light on their evaporation and possible information loss  \cite{AOSRBH}. Moreover, it has been argued that quantum phenomena can leave observable traces in gravitational waves emitted by black holes, via tidal heating, echoes, or by modifying the ringdown of perturbed black holes (see e.g. Refs. \cite{AguGW,Corral}). Another important problem (not unrelated to the previous ones) is the quantum fate of the essential singularities that are inherent to the interior of black holes in GR. It is expected that these singularities can be resolved by quantum processes, and in particular by a loop quantization. This resolution by loop techniques would be similar to the removal of the big bang singularity that occurs in LQC \cite{A&S,APS}.

The interest in this question was recently revitalized when Ashtekar, Olmedo, and Singh (AOS) introduced an effective model to describe LQC modifications to a Kruskal geometry \cite{AOS,AOS2,AO}. This model satisfactorily includes small quantum corrections near the horizon of the black hole for large mass values, predicts upper bounds for the curvature invariants that are mass independent, and truly resolves the classical singularity, which is substituted by a transition surface that connects a trapped region with an anti-trapped one. Nonetheless, the model has received different kinds of criticisms \cite{Bouh}, e.g. questioning the role of general covariance \cite{LBH9,BojN} (see also the recent proposal of Ref. \cite{GPN}), or remarking the fact that its equations of motion seem in conflict with a straightforward derivation from a Hamiltonian formalism \cite{Nor1,GQM,GQM2}. Focusing our attention on this last issue, we notice that it poses an important obstacle on the route to quantization. The absence of a well defined Hamiltonian formulation prevents a canonical quantization (even by nonperturbative techniques). Furthermore, since the appealing physical properties of the AOS model arise as a consequence of its dynamics, changing its equations of motion is not a desirable alternative. To circumvent this problem, the proposers of the model considered an extended phase space formulation \cite{AOS2}, in which the parameters that encode the quantum effects are treated as variables that are constrained to be functions of the black hole mass. This mass is a phase space function that remains constant throughout the evolution, i.e. on dynamical solutions. The extended formulation includes the aforementioned parameters as configuration degrees of freedom, introducing suitable momenta for them. Then, in order to preserve the number of degrees of freedom that are physical, one imposes as constraints the relations between those parameters and the phase space functions that determine them in terms of the black hole mass. The Hamiltonian dynamics of this extended formulation reproduces the original equations of motion of the AOS model. Recent studies have confirmed that such dynamics is indeed maintained under reduction of the system \cite{BH_GAB}. However, an important subtlety appears in the reduction process that had been initially overlooked \cite{AOS2}: the reduced phase space has a symplectic structure that is not equivalent to that of the original AOS model in GR \cite{BH_GAB}. Actually, the nonstandard symplectic structure of the reduced space is what ensures the correct derivation of the dynamics, by changing the Poisson brackets into nonequivalent Dirac brackets \cite{DN}. This intricate reduced symplectic structure makes practically unviable a direct quantization of the AOS model using loop techniques. On the contrary, the formalism in the extended phase space emerges as a promising candidate to carry out the quantization of the interior region of the black hole by employing nonperturbative canonical methods.

The aim of this work is to achieve a complete quantization of the extended phase space version of the AOS model. We will follow the strategy outlined in Ref. \cite{BH_GAB}, where a formal quantum analysis was first presented. However, we will modify the density weight of the Hamiltonian constraint analyzed in that reference, and discuss in full detail the construction of the quantum counterpart of this constraint and the specification of its solutions. Specifically, we will adopt a densitization of the Hamiltonian constraint like the one that is usually employed in LQG \cite{A&L}, and which has been suggested in other scenarios of LQC \cite{Prescrip_G}. In this way, we will avoid recurring to the inverse of certain geometric operator in the construction of such constraint (see Ref. \cite{BH_GAB}). This inverse complicates the determination of a suitable domain of definition for the constraint operator and of a conveniently large set of dual states to search for its solutions. Actually, the densitization we will use has previously been employed in other works that have considered the quantization of nonrotating black holes in LQC \cite{BH_Con,BH_Cong}. In addition to the fact that, in principle, those works do not contemplate an extension of the phase space of the model, an important difference with respect to them is that we will manage to consider a single kinematic Hilbert space for the geometric part of the system in the construction of the Hamiltonian constraint operator, obtaining the corresponding space for the full extended system as a tensor product. 
This will be possible by rescaling the geometric variables in such a way that the action of the relevant operators becomes independent of the value of the parameters that regulate the quantum effects \cite{BH_GAB}.

We will define the (geometric) Hamiltonian constraint operator along lines similar to those that have been thoroughly discussed in LQC. The rest of constraints, restricting the quantum parameters of the extension, are simple to impose. Focusing our attention on the nontrivial Hamiltonian constraint, we will perform a spectral analysis of the geometric operators that compose it. Choosing a suitable dense set in (the geometric part of) the kinematic Hilbert space as the common domain of definition for those operators and seeking for solutions to the constraint in its algebraic dual, we will derive an algorithm to compute them in closed form for any value of the regularization parameters. This algorithm specifies the wavefunction that describes the radial part of the metric [associated to a canonical pair $(b,p_b)$] in terms of only two initial data. We will examine the asymptotic behavior of such solutions for large rescaled triad variables, asymptotics that is commonly understood as a Wheeler-DeWitt (WDW) limit. This analysis will not only assist us in solving the issue of how to deal with the two initial values appearing in the resolution algorithm, but it also will offer a means to determine the possibly divergent part of our solutions with respect to the kinematic norm. With all this information, we will be in adequate conditions to specify the set of states annihilated by all the constraints in our extended phase space formalism and, moreover, endow it with a suitable inner product. This product will provide us with the physical Hilbert space, completing the quantization.

The paper is structured as follows. In Sec. \ref{sec: II} we introduce the classical formalism and the kinematic Hilbert space for the geometric degrees of freedom. Sec. \ref{sec: III} contains a discussion of the extended phase space framework analyzed in Ref. \cite{BH_GAB} and of the quantization of its constraints for our choice of densitization for the lapse function. In Sec. \ref{sec: IV} we calculate the explicit expressions of the states annihilated by the Hamiltonian constraint. We also study their asymptotic behavior when the radial geometric variable becomes large. The construction of the physical states is carried out in Sec. \ref{sec: V}. Finally, we summarize our results and present some further discussion in Sec \ref{sec: VI}. Two appendices are added, which provide details about the spectral analysis of geometric operators and about the WDW limit. Throughout the article, we employ Planck units, with the speed of light and the Planck and Newton constants equal to one.

\section{Cosmological model for the interior geometry \label{sec: II}}

In this section, we succintly describe the classical formulation of the model and its kinematic representation along the lines of LQC. More information about the interior geometry and the construction behind the AOS model can be found in Refs. \cite{K&S,E.Weber,A&B,B&V,AOS, AOS2}.

The interior region of a nonrotating black hole can be modeled using a Kantowski-Sachs metric \cite{K&S, E.Weber}, with an appropriate choice of coordinates \cite{A&B}. Its line element is given by
\begin{equation}
ds^2 = -N(\tau)^2\text{d}\tau^2 + \frac{p_b^2(\tau)}{L^2_o|p_c(\tau)|} \text{d}x^2 + |p_c(\tau)| \text{d}\Omega_2^2,
\end{equation}
where $N$ is the lapse function, $\theta \in \left[0,\pi\right)$, $\phi \in \left[0,2\pi\right)$ and $x\in [0,L_o]$, with $L_o$ being a fiducial length introduced to avoid infrared divergences. In LQC, the phase space can be described using as coordinates the geometric degrees of freedom of the Ashtekar-Barbero connection and of the densitized triad. Using the spatial homogeneity of the model and a suitable gauge for the internal SU(2) freedom \cite{A&B}, we can express the connection and triad variables as
\begin{eqnarray}
\label{eq: II_HKS}
E^{\alpha}_i \partial_{\alpha} &=&\; \delta^3_i p_c \text{sin}\theta \partial_x + \delta^2_i \frac{p_b}{L_o} \text{sin}\theta \partial_{\theta} - \delta^1_i \frac{p_b}{L_o} \partial_{\phi},\\
\label{eq: II_HKS_A}
A^i_{\alpha} \text{d}x^{\alpha} &=&\; \delta^i_3 \frac{c}{L_o}\text{d}x + \delta^i_2b\text{d}\theta - \delta^i_1b\text{sin}\theta \text{d}\phi + \delta^i_3 \text{cos}\theta\text{d}\phi.
\end{eqnarray}
In these formulas, the letters $\alpha$ and $i$ respectively represent spatial and internal SU(2) indices. We call $\Gamma_{\text{KS}}$ the phase space of this geometric model, with a symplectic structure given by the canonical pairs $\{ b,p_b \} = \gamma$, $\{ c,p_c \} = 2\gamma$ , where $\gamma$ is the Immirzi parameter \cite{A&L, Thiemann}. The only nontrivial constraint that remains to be satisfied is the Hamiltonian constraint, that takes the form
\begin{gather}
H_{\text{KS}}\left[N\right] = N L_o\frac{b}{\gamma\sqrt{|p_c|}}\left(O^{\text{KS}}_b - O^{\text{KS}}_c\right),
\end{gather}
where 
\begin{gather}
O^{\text{KS}}_b = -\frac{p_b}{2\gamma L_o} \left( b + \frac{\gamma^2}{b} \right), \qquad O^{\text{KS}}_c = \frac{cp_c}{\gamma L_o}.
\end{gather}
We will call these phase space functions the partial Hamiltonians, since they separately generate the dynamics of each of the two geometric canonical pairs of the system. They are both constants of motion that become equal in norm to the ADM mass of the black hole when we are \textit{on-shell}. 
Selecting a lapse function of the form $N_{\text{dec}}=\gamma \sqrt{|p_c|}/b$ greatly simplifies the corresponding equations of motion, since the constraint decouples into two independent sectors.
	
The fundamental variables employed in LQG are the holonomies of the connection and the fluxes of the densitized triad \cite{A&L, Thiemann}. Restricting our discussion to a Kantowski-Sachs spacetime, the holonomies along the edges in the $\theta$ and $x$ directions contain all the relevant information about the connection. The holonomy matrix elements are determined by complex exponentials of the form $\mathcal{N}_{\mu_j} = e^{ij\mu_j/2}$, where $\mu_j \in \mathbb{R}$ is a coordinate length parameter for the considered edge, with $j=b$ or $c$. Fluxes over surfaces, bounded by edges in either coordinate direction, are defined in terms of the variables $p_j$. Describing the configuration space in terms of the holonomy variables leads to considering two copies of the algebra of almost periodic functions. Homogeneous and isotropic systems in LQC are described with just one copy of such algebra, and the loop quantization procedure of these models is well established \cite{ABL,Vel}. Following similar techniques in the present case, the kinematic Hilbert space is constructed in the densitized triad representation by completing each copy of the algebra with respect to the discrete product and taking the tensor product of the two individual Hilbert spaces. We will call $\mathcal{H}_{\text{LQC}}^{\text{kin}}$ this kinematic space. A basis of each individual space is provided by the eigenstates $|\mu_j\rangle$, with $\mu_j \in \mathbb{R}$ indicating the eigenvalue of $p_j$. These states are normalized to the unity. 

In the next section we will see how the construction of the Hamiltonian operator involves certain regularization parameters $\delta_j$. As shown in Ref. \cite{BH_GAB}, when passing to the extended phase formulation, in which such parameters are treated as new variables, it is most convenient to introduce the rescaling $\tilde{p}_j = p_j/\delta_j$. Accordingly, it is useful to relabel our basis in terms of the eigenvalue of $\tilde{p}_j$, i.e. $\tilde{\mu}_j = \mu_j/\delta_j$. Note that the action of the holonomy elements along edges of coordinate length $\delta_j$ are then independent of the value of $\delta_j$. Explicitly, we have after the rescaling that 
\begin{equation}
\hat{\mathcal{N}}_{\delta_j} |\tilde{\mu}_j \rangle = |\tilde{\mu}_j + 1 \rangle.
\end{equation} 
On the other hand, fluxes are given by 
\begin{equation}\label{ps}
\hat{p}_b |\tilde{\mu}_b\rangle = \frac{\gamma \tilde{\mu}_b \delta_b}{2} |\tilde{\mu}_b\rangle, 
\qquad \hat{p}_c |\tilde{\mu}_c\rangle = \gamma \tilde{\mu}_c \delta_c |\tilde{\mu}_c\rangle.
\end{equation} 
Of course, the corresponding action of $\hat{\tilde{p}}_j$ is independent of $\delta_j$, by construction. Hence, if we define all relevant geometric operators
in terms of $\hat{\mathcal{N}}_{\delta_j}$ and $\hat{\tilde{p}}_j$, their action on our kinematic Hilbert space of the black hole interior becomes independent of the $\delta$-parameters of the model.

\section{Constraint operator and extended model\label{sec: III}} 

After introducing our kinematic representation, our next step is to promote the Hamiltonian constraint to an operator. Let us recall that the representation of the holonomy-flux algebra is discrete and, as a consequence, noncontinuous, so that the connection variables cannot be defined as derivatives of our holonomy operators. This problem, which also arises in full LQG, is solved by means of an established regularization procedure \cite{Thiemann_Reg}. In brief, the procedure assumes a minimally small but nonzero value for the area enclosed by the basic holonomy circuits, formed by edges with lengths that are fixed by certain regularization parameters \cite{APSIm}. The curvature-holonomy relation is then truncated at dominant order in the small edge size. For the considered Kantowski-Sachs spacetime, the parameters are denoted by $\delta_j$. This procedure leads to a regularized Hamiltonian that is often understood as effective and is supposed to include quantum corrections with respect to the Kantowski-Sachs Hamiltonian, incorporated via its dependence on the $\delta$-parameters:
\begin{gather}
\label{eq: III_Heff}
H^{\text{eff}}_{\text{KS}}\left[N\right] = N \frac{L_o}{\gamma\sqrt{|p_c|}} \frac{\text{sin}(\delta_b b)}{\delta_b} (O_b - O_c),
\end{gather}
where
\begin{gather}
O_b = -\frac{p_b}{2\gamma L_o} \left[ \frac{\text{sin}(\delta_b b)}{\delta_b} + \frac{\gamma^2\delta_b}{\text{sin}(\delta_b b)} \right], \qquad
O_c = \frac{\text{sin}(\delta_c c)}{\gamma L_o \delta_c}p_c,
\end{gather}
are the effective partial Hamiltonians \cite{AOS,AOS2,BH_GAB}. Following the same arguments as in the classical model, the constant of motion $O_c$ is identified with the black hole mass of the effective solutions, in absolute value. Notice that, in the limit $\delta_j \rightarrow 0$, we recover the genuinely classical Hamiltonian given in Eq. \eqref{eq: II_HKS}. 

Several proposals have been suggested to fix the value of the $\delta$-parameters, that go from the possibility of keeping them as constant numbers \cite{LBH1,A&B,LBH3} to letting them be functions on phase space that change \cite{B&V,Chiou} or remain constant \cite{LBH5,JPS} along dynamical trajectories. In this paper, our emphasis is on the scheme proposed by the authors of the AOS model. In this case, the parameters are determined by area arguments on a transition surface where the metric function $p_c$ finds its minimum in the model \cite{AOS,AOS2}. This imposition fixes $\delta_j$ as functions of only the effective partial Hamiltonians on-shell such that, when the value $m$ of these Hamiltonians is large (in norm), they satisfy 
\begin{gather}\label{deltaconsi}
\delta_b = \left( \frac{\sqrt{\Delta}}{\sqrt{2\pi}\gamma^2m}\right)^{1/3}, \qquad \delta_c = \frac{1}{2 L_o}\left( \frac{\gamma\Delta^2}{4\pi^2m}\right)^{1/3}.
\end{gather}
Here, $\Delta$ is the area gap of LQG, i.e., the minimum nonvanishing value allowed for the physical area by the spectrum of the area operator \cite{A&L,APS}. As we have mentioned, the justification for these expressions is based on considerations of this minimal physical area on a transition surface that replaces the essential singularity of GR in the AOS model \cite{AOS,AOS2}. It is worth emphasizing that these considerations lead to regularization parameters that are given by functions of the partial Hamiltonians and therefore become, strictly speaking, functions on the phase space of the system. This dependence of the parameters on the geometry is the parallel for black holes of the procedure advocated in LQC for the quantization of homogeneous cosmologies, which has proven successful in a variety of scenarios \cite{A&S,APSIm,BI}.

As we commented in the Introduction, recent studies \cite{BH_GAB} have shown that, in order to obtain a consistent Hamiltonian formulation of the AOS model and base the quantization process on it, it is necessary to extend the phase space of the Kantowski-Sachs cosmology. The extension consists in including the two parameters $\delta_j$ as configuration variables into a more general space, that is endowed with a canonical symplectic structure to transform it into a phase space. To preserve the number of physical degrees of freedom, two constraints are introduced in this extended system, which impose that the parameters satisfy on-shell the functional relations \eqref{deltaconsi} that result from the existence of a minimal physical area in LQG \cite{AOS,AOS2}. We will call $\Gamma_{\text{ext}}$ this extended phase space. 

More concretely, in addition to the pairs $(j,p_j)$, $\Gamma_{\text{ext}}$ contains two other canonical pairs $(\delta_j,p_{\delta_j} )$ corresponding to the $\delta$-parameters \cite{AOS2}. The availability of a Hamiltonian formalism with manageable canonical structure allows the passage to the quantum theory following loop techniques \cite{BH_GAB}. The total Hamiltonian of the extended model becomes
\begin{gather}
\label{eq: II_Hext^eff}
H^{\text{eff}}_{\text{ext}} = \frac{N L_o}{\gamma\sqrt{|p_c|}} \frac{\text{sin}(\delta_b b)}{\delta_b} \left(O_b - O_c\right) + \lambda_b\Psi_b + \lambda_c\Psi_c, \qquad \Psi_j = K_j(O_b,O_c) - \delta_j,
\end{gather}
where $\Psi_j$ are the constraints that impose the relation between the $\delta$-parameters and the partial Hamiltonians, namely, the parameters must coincide with functions on phase space that reproduce the dependence on the black hole mass \eqref{deltaconsi} (for large $m$ in norm) when the partial Hamiltonians are evaluated on-shell. We assume that $K_j(O_b,O_c)$ are sufficiently smooth functions that satisfy \cite{BH_GAB}
\begin{gather}
\label{eq: II_Ktilde}
K_b(m,m)=\tilde{K}_b(m)  \longrightarrow \left( \frac{\sqrt{\Delta}}{\sqrt{2\pi}\gamma^2 m} \right)^{1/3}, \qquad
K_c(m,m)=\tilde{K}_c(m) \longrightarrow \frac{1}{2 L_o}\left( \frac{\gamma\Delta^2}{4\pi^2 m} \right)^{1/3},
\end{gather}
when the absolute value of $m$ is large. Note that, on-shell, the constraints $\Psi_j$ amount to imposing that $\delta_j$ equal the functions $\tilde{K}_j$ evaluated on the partial Hamiltonian $O_c$. A simple calculation (along the lines explained in Appendix A of Ref. \cite{AOS2}) proves that the three constraints that appear in Eq. \eqref{eq: II_Hext^eff} are first-class in the sense of Dirac \cite{DN}.

The above total Hamiltonian indeed reproduces the dynamics of Eq. \eqref{eq: III_Heff} if the Lagrange multipliers $\lambda_j$ vanish and the evolution of the geometric variables takes place on the constraint surface defined by $\Psi_j$ \cite{DN}. It is worth commenting that this vanishing of $\lambda_j$ can be enforced with a suitable gauge fixing on $\Gamma_{\text{ext}}$ that leads to this condition for consistency. This gauge fixing would result in a reduced system with a symplectic structure for the geometric variables that differs from that of the original Kantowski-Sachs cosmology. A detailed analysis of this reduction can be found in Ref. \cite{BH_GAB}. In summary, if one wants to maintain the effective dynamics of the AOS model, and therefore its geometric solutions with appealing physical properties, the only manner to stay in the original phase space is to modify the symplectic structure with respect to that found in GR. But the modified symplectic structure is then so intrincate that the quantization of the system is not viable. In particular, one cannot adopt a nonperturbative canonical approach based on LQG to quantize the model in that way.

The extension of the phase space allows one to avoid a complicated symplectic structure for the phase space of the geometry in terms of connection and triad variables. Therefore, it offers a good road to the quantization of the black hole interior. An extended kinematic representation can be constructed by quantizing the geometric degrees of freedom with loop techniques, as shown in Sec. \ref{sec: II}, while the $\delta$-parameters are quantized using a continuous Schrödinger representation, with Hilbert spaces defined as $L^2(\mathbb{R},\text{d}\delta_j)$, with the corresponding Lebesgue measure. This choice is justified by the fact that one expects LQG phenomena to be most important for the genuine gravitational degrees of freedom. The $\delta$-parameters, on the other hand, are real scalar quantities off-shell, and not components of densitized triads or connections. These parameters are only constrained by the relations that determine them as functions of the geometric variables on-shell. Hence, we choose to adopt a continuous representation for them, in parallel with similar works in which the system contains a minimally coupled scalar field \cite{ABL}, also described quantum mechanically in a Schrödinger representation.  The resulting kinematic Hilbert space is the tensor product $\mathcal{H}_{\text{ext}}^{\text{kin}} = \mathcal{H}_{\text{LQC}}^{\text{kin}} \otimes L^2(\mathbb{R},\text{d}\delta_b) \otimes L^2(\mathbb{R},\text{d}\delta_c)$. A convenient basis is given by the states $|\tilde{\mu}_b,\tilde{\mu}_c,\delta_b,\delta_c\rangle$ obtained as products of the eigenstates of the rescaled triad variables (with $|\tilde{\mu}_j\rangle$ normalized to the Kronecker delta) and eigenstates of the $\delta$-parameters (with $|\delta_j\rangle$ normalized to the Dirac delta).

Finding a representation for Eq. \eqref{eq: II_Hext^eff} is now within reach. Let us first specify the densitization of that constraint. Rather than employing the same one as in Ref. \cite{BH_GAB}, it proves better to use a constraint of unit density weight proportional to the volume of the spatial sections. This densitization parallels the standard one in LQG \cite{A&L}. It has also been employed in different studies of homogeneous and isotropic cosmologies \cite{Prescrip_G,hyb-review}, and has been applied in recent works on nonrotating black holes \cite{BH_Con,BH_Cong}. This alternative lapse not only solves some of the problems encountered in Ref. \cite{BH_GAB}, related with the inversion of geometric operators, but in addition allows for the definition of the operator representing the constraint \eqref{eq: II_Hext^eff} in terms of its action on the basis elements $|\tilde{\mu}_b,\tilde{\mu}_c,\delta_b,\delta_c\rangle$. 

Taking $N = \sqrt{|p_c|}p_b \underline{N}$, with $\underline{N}$ the lapse function of the densitized constraint according to our previous comments, and adopting a factor ordering known as the MMO prescription \cite{MMO}, which has proven advantageous in the study of homogeneous and isotropic LQC \cite{hyb-review}, we then obtain the following operator representation for the Kantowski-Sachs contribution to the total Hamiltonian of the extended system (up to multiplication by $\underline{N}$):
\begin{gather}
\label{eq: II_HLQC}
\hat{H}_{\text{LQC}} = -\frac{1}{2}\left( \hat{\Omega}_b^2 + \delta_b^2 \hat{\tilde{p}}_b^2 + 2\hat{\Omega}_b\hat{\Omega}_c \right).
\end{gather}
The operator $\hat{\Omega}_j$ represents the product $\tilde{p}_j \sin{(\delta_j j)}/\gamma$. Calling $\widehat{\sin{(\delta_ j j)}}=(\hat{\mathcal{N}}_{2\delta_j}-\hat{\mathcal{N}}_{-2\delta_j})/(2i)$, we explicitly have \cite{MMO,BH_GAB}
\begin{eqnarray}\label{Omegab}
\hat{\Omega}_j=\frac{1}{2\gamma}|\hat{\tilde{p}}_j|^{1/2}\left[\widehat{\sin(\delta_j j)}\widehat{\text{sign}(\tilde{p}_j)}+\widehat{\text{sign}(\tilde{p}_j)}\widehat{\sin(\delta_j j)}\right]|\hat{\tilde{p}}_j|^{1/2}.
\end{eqnarray}
Note that the action of the two operators $\hat{\Omega}_j$ is $\delta$-independent. Finally, $\widehat{\text{sign}}$ is the sign operator. In the following, we will refer to the operator \eqref{eq: II_HLQC} as the Hamiltonian constraint operator, or quantum Hamiltonian constraint.

In the above definition of $\hat{H}_{\text{LQC}}$,  we have used that the $\delta$-parameters act as multiplication operators in the representation that we have adopted. This is in consonance with their consideration as configuration variables of the extended phase space. At this stage, these variables are real ones, similar to the position variables of ordinary Quantum Mechanics. We still have to impose on them the constraints that relate them to the black hole mass on-shell and that compensate the extension of the physical phase space that we have performed. The imposition in the quantum theory of such remaining constraints, namely $\Psi_j$, is achieved by implementing the spectral theorem to convert the functions $K_j$ on-shell, i.e. $\tilde{K}_j$, into functions of the operator that represents the partial Hamiltonian $O_c$, in agreement with our previous discussion. With our representation, this partial Hamiltonian is given by $\hat{\Omega}_c/L_o$. Let us now turn our attention to the properties of the geometric operators involved in the above quantum Hamiltonian constraint. In fact, they are very similar to the operators for LQC studied in Refs. \cite{MMO,BI_MGT}.

\subsection{Analysis of the operators $\hat{\Omega}_{j}^2$ and $\hat{\Omega}_{j}$}

Examining the definition of $\hat{\Omega}_{j}^2$, we see that it is a difference operator relating states that differ in 4 units in the label $\tilde{\mu}_{j}$. This operator is (essentially) self-adjoint, with an absolutely continuous and nondegenerate spectrum. In fact, by construction, it is a positive operator. A proof of these statements can be found in Appendix \ref{sec: App_SA}. As we have commented, the operator only relates basis elements with label $\tilde{\mu}_{j}$ in any semilattice of the form $^{(4)}\mathcal{L}_{\tilde{\varepsilon}_{j}}^{\pm} = \{ \pm(\tilde{\varepsilon}_{j} + 4n) :  n\in\mathbb{N} \}$, with $\tilde{\varepsilon}_{j} \in \left(0,4\right]$. The fact that one can restrict its action to semilattices is a consequence of the MMO factor ordering, that guarantees that triad orientations are preserved. Let us call $^{(4)}\mathcal{H}_{\tilde{\varepsilon}_{j}}^{\pm}$ the Hilbert space obtained by completing the linear span $^{(4)}\text{Cyl}_{\tilde{\varepsilon}_{j}}^{\pm} = \text{span} \{ |\tilde{\mu}_{j}\rangle : \tilde{\mu}_{j} \in\! ^{(4)}\mathcal{L}_{\tilde{\varepsilon}_{j}}^{\pm} \}$ with respect to the discrete product. These Hilbert spaces do not get mixed under the action of $\hat{\Omega}_{j}^2$. The parameter $\tilde{\varepsilon}_j$ is the the minimum value of $|\tilde{\mu}_{j}|$ in the considered space.

If we define $\omega_b = 1/8$ and $\omega_c = 1/4$, the action of $\hat{\Omega}_{j}^2$ on the eigenstates of $\tilde{p}_j$ can be expressed as 
\begin{eqnarray}
\label{eq: III_O^2}
\hat{\Omega}_{j}^2 |\tilde{\mu}_{j}\rangle &=& - \omega_j^2 \left\{ f_{+}(\tilde{\mu}_j + 2)f_{+}(\tilde{\mu}_j)|\tilde{\mu}_j + 4\rangle  - [ f_{+}^2(\tilde{\mu}_j) + f_{-}^2(\tilde{\mu}_j)] |\tilde{\mu}_j\rangle + f_{-}(\tilde{\mu}_j - 2)f_{-}(\tilde{\mu}_j)|\tilde{\mu}_j - 4\rangle \right\},\\
f_{\pm}(\tilde{\mu}) &=& |\tilde{\mu} \pm 2|^{1/2}[\text{sign}(\tilde{\mu} \pm 2) + \text{sign}(\tilde{\mu})]|\tilde{\mu}|^{1/2}. \label{fpm}
\end{eqnarray}
Without loss of generality, we can simplify our analysis by focusing on one of the invariant Hilbert spaces with positive orientation. It is possible to show that the generalized eigenstates $| e_{m_{j}^2}^{\tilde{\varepsilon}_{j}} \rangle$, for any positive eigenvalue $m_{j}^2$, are completely determined by a single initial datum for their coefficients at $\tilde{\varepsilon}_j$ \cite{MMO}:  
\begin{equation}
| e_{m_{j}^2}^{\tilde{\varepsilon}_{j}} \rangle = \sum_{\tilde{\mu}_j \in^{(4)}\! \mathcal{L}^{+}_{\tilde{\varepsilon}_{j}}} e_{m_{j}^2}^{\tilde{\varepsilon}_{j}}(\tilde{\mu}_j) | \tilde{\mu}_j \rangle, \qquad
e_{m_{j}^2}^{\tilde{\varepsilon}_{j}}(\tilde{\varepsilon}_{j} + 4n) =  \left[S_{\tilde{\varepsilon}_{j}}(0,2n) + \frac{F(\tilde{\varepsilon}_{j})S_{\tilde{\varepsilon}_{j}}(1,2n)}{G_{|m_j|}(\tilde{\varepsilon}_{j} -2)}\right] e_{m_{j}^2}^{\tilde{\varepsilon}_{j}}(\tilde{\varepsilon}_{j}). 
\end{equation}
Here, $F(\tilde{\mu}) = f_{-}(\tilde{\mu})/f_{+}(\tilde{\mu})$, $G_{|m_j|}(\tilde{\mu}) = -i|m_j|/[\omega_jf_{+}(\tilde{\mu})]$ and
\begin{equation}\label{S}
S_{\tilde{\varepsilon}_j}(u,v)=\sum_{O(u\rightarrow v)}\left[\prod_{\{r_u\}}F(\tilde{\varepsilon}_j+2 r_u +2)\prod_{\{s_u\}}
G_{|m_j|}(\tilde{\varepsilon}_j+2 s_u ) \right],
\end{equation}
where $O(u\rightarrow v)$ is the set of all possible jumps of one or two units from $u$ to $v$, with $\{s_u\}$ and $\{r_u\}$ the corresponding subsets of integers followed by a jump of one or two unit steps, respectively \cite{MMO}. Our compact notation makes use of the fact that $F(\tilde{\varepsilon}_j)=0$ for all $\tilde{\varepsilon}_j\leq 2$. The coefficients of these eigenstates can then be taken all real. We fix the initial datum by letting it be positive and normalizing the eigenstates so that $\langle e_{m_{j}^2}^{\tilde{\varepsilon}_{j}} | e_{\tilde{m}_{j}^2}^{\tilde{\varepsilon}_{j}} \rangle = \delta(m_{j}^2 - \tilde{m}_{j}^2)$.

In turn, $\hat{\Omega}_{j}$ is a difference operator that connects states separated by 2 units in the label $\tilde{\mu}_{j}$. It is (essentially) self-adjoint, with absolutely continuous and nondegenerate spectrum equal to the real line. These properties are also discussed in Appendix \ref{sec: App_SA}. The action of the operator on the eigenstates of $\tilde{p}_j$ is given by
\begin{gather}
\hat{\Omega}_j |\tilde{\mu}_j\rangle = i \omega_j \left[ f_{-}(\tilde{\mu}_j) |\tilde{\mu}_j - 2\rangle - f_{+}(\tilde{\mu}_j) |\tilde{\mu}_j + 2 \rangle\right].
\end{gather}
It preserves the triad orientation and, furthermore, leaves invariant the Hilbert spaces $^{(2)}\mathcal{H}_{\tilde{\epsilon}_{j}}^{\pm} =\, ^{(4)}{\mathcal{H}}_{\tilde{\epsilon}_{j}}^{\pm} \oplus\, ^{(4)}{\mathcal{H}}_{\tilde{\epsilon}_{j} + 2}^{\pm}$, with $\tilde{\epsilon}_{j} \in (0,2]$. Actually, these spaces are not mixed by the repeated action of the Hamiltonian constraint operator, and in this sense we can consider that they provide superselection sectors of the model \cite{A&L,A&S,MMO}. We restrict our considerations, for simplicity, to a Hilbert space with positive orientation. The generalized eigenstates $| e^{\tilde{\epsilon}_j}_{m_j} \rangle$ of $\hat{\Omega}_{j}$, with eigenvalues $m_{j}\in\mathbb{R}$, can be related to eigenstates of the squared operator in the two complementary semilattices of points separated by four units that form the support of the superselection sector. One can show that \cite{BI_MGT} 
\begin{equation}
| e^{\tilde{\epsilon}_j}_{m_j} \rangle = \sqrt{|\smash{m_j}|}[| e^{\tilde{\epsilon}_j}_{m_j^2} \rangle \oplus i\;\text{sign}(-m_j) | e^{\tilde{\epsilon}_j + 2}_{m_j^2} \rangle],
\end{equation} 
for all $m_j \neq 0$. The normalization of these states is such that $\langle e_{m_{j}}^{\tilde{\epsilon}_{j}} | e_{\tilde{m}_{j}}^{\tilde{\epsilon}_{j}} \rangle = \delta(m_{j} - \tilde{m}_{j})$.

\subsection{WDW limit of ${\Omega}_{j}^2$ and ${\Omega}_{j}$\label{ssec: WDW}}

The normalization of our generalized eigenstates in terms of Dirac deltas can be achieved by means of an asymptotic analysis of their coefficients when $\tilde{\mu}_j$ takes large (absolute) values. In this limit, in which the discreteness of the support of the superselection sectors becomes comparatively small, we expect to recover a quantum behavior proper to wavefunctions of the WDW theory in geometrodynamics. Once more, we focus our analysis on superselection sectors with positive orientation.

For ${\hat\Omega}_{j}^2$, the limit leads to the differential operator $\underline{\Omega}_j^2 = -16\omega_j^2[ 1 + 2\tilde{\mu}_{j}\partial_{\tilde{\mu}_{j}}]^2$. Details can be found in Appendix \ref{sec: App_WDW}. Therefore, the real eigenstates of ${\hat\Omega}_{j}^2$ tend to 
\begin{eqnarray}
\underline{e}_{m_{j}^2}^{\tilde{\varepsilon}_{j}}(\tilde{\mu}_{j}) &=& R [ e^{i\varphi_j } \underline{e}^{+}_{|m_j|}(\tilde{\mu}_{j}) + e^{-i\varphi_j } \underline{e}^{-}_{|m_j|}(\tilde{\mu}_{j}) ],\nonumber\\
\underline{e}^{\pm}_{|m_j|}(\tilde{\mu}_{j}) &=& \frac{1}{\sqrt{16\pi\omega_j \tilde{\mu}_{j}}} \text{exp}\left\{ \mp i\frac{|m_j|}{8\omega_j}\ln \tilde{\mu}_{j} \right\},
\end{eqnarray} 
where we recall that $\tilde{\mu}_j>0$ in our superselection sector of positive orientation, and $\varphi_j\in S^1$ and $R>0$ are constants. The latter is fixed to be equal to $2$ by requiring a Dirac-delta normalization and using the relation between the LQC and WDW norms \cite{Prescrip_G,K&P}. The phase $\varphi_j$, on the other hand, displays a well known dependence on $m_j$ and $\tilde{\varepsilon}_j$ that is not needed for the purposes of this work \cite{MMO,K&P}.

In the case of ${\hat\Omega}_{j}$, one would be tempted to think that the limit is a first order differential operator that coincides with the square root of $\underline{\Omega}_{j}^2$ (up to sign). However, the reality is that this operator does not admit a smooth limit in the whole superselection sector $^{(2)}\mathcal{H}_{\tilde{\epsilon}_{j}}^{+}$. The reason is that its generalized eigenfunctions exhibit a very rapid oscillation in phase whenever they vary from one to the other of the two complementary semilattices of points separated by four units that compose the support of the states. However, they display a smooth asymptotic behavior for each of these two semilattices, adopting therefore two different limits, one for $ ^{(4)}\mathcal{L}^{+}_{\tilde{\epsilon}_{j}}$ and another for $ ^{(4)}\mathcal{L}^{+}_{\tilde{\epsilon}_{j} + 2}$. They are given by
\begin{equation}
\sqrt{R}[ e^{i\varphi_j} \underline{e}^{+}_{|m_j|}+  e^{-i\varphi_j} \underline{e}^{-}_{|m_j|}] \Upsilon_{m_j} , 
\end{equation}
where $\Upsilon_{m_j}=1$ when $\tilde{\mu}_{j} \in\! ^{(4)}\mathcal{L}^{+}_{\tilde{\epsilon}_{j}}$ and $\Upsilon_{m_j}=-i\mathrm{sign}(m_j)$ when $\tilde{\mu}_{j} \in\! ^{(4)}\mathcal{L}^{+}_{\tilde{\epsilon}_{j}+2}$ \cite{BID_MGT}.

\section{Quantum Hamiltonian constraint on the geometry\label{sec: IV}} 

After studying the operators that appear in Eq. \eqref{eq: II_HLQC}, we are ready to analyze this Hamiltonian constraint operator as a whole. From our definition of all the involved geometric operators, we immediately see that the linear span of the eigenstates of the rescaled triad variables $\tilde{p}_j$ is an acceptable domain of definition for this quantum Hamiltonian constraint for any given value of $\delta_b$ or, equivalently, on any generalized eigenspace of $\hat{\delta}_b$. In the following, we concentrate our analysis on a superselection sector for the geometry of the type $^{(2)}\!\mathcal{H}_{\tilde{\epsilon}_{b}}^{+}\otimes^{(2)}\!\mathcal{H}_{\tilde{\epsilon}_{c}}^{+}$. Moreover, since $\hat{\Omega}_c$ commutes with the constraint, we can further focus the discussion on any generalized eigenstate of this operator, with eigenvalue $m_c$, which is directly related to the mass $|m|$ of the black hole by $m_c = m L_o $. Then, the quantum Hamiltonian constraint can be reexpressed as the following equation on any quantum state of the geometry of the $b$-sector, $|\psi^{\tilde{\epsilon}_b}_{\delta_b}\rangle \in ^{(2)}\!\mathcal{H}_{\tilde{\epsilon}_{b}}^{+}$:
\begin{gather}\label{Qmass}
\hat{\mathcal{Q}}_b\left(m_c\right) |\psi^{\tilde{\epsilon}_b}_{\delta_b}\rangle  = \left[ (\hat{\Omega}_b + m_c )^2 + \delta_b^2\hat{\tilde{p}}_b^2\right] |\psi^{\tilde{\epsilon}_b}_{\delta_b}\rangle = m_c^2 |\psi^{\tilde{\epsilon}_b}_{\delta_b}\rangle .
\end{gather}
The operator $\hat{\mathcal{Q}}_b(a)$ can be proven to have a discrete spectrum, $\sigma(a)$, and each of its eigenvalues $\rho\in\sigma(a)$ describes an analytic curve as a function of the parameter $a$. These issues are discussed in Appendix \ref{sec: App_SA}. Actually, a very similar operator was considered in Ref. \cite{BH_Cong}, demonstrating that the corresponding spectrum is discrete. Hence, solutions to the Hamiltonian constraint equation above can be obtained as eigenstates of $\hat{\mathcal{Q}}_b(a)$ with eigenvalues that satisfy the condition $\rho = a^2$. The intersection between the analytic curves of the eigenvalues of $\hat{\mathcal{Q}}_b(a)$ and the also analytic curve $a^2$ only occur at isolated points, resulting in a discrete set of possible solutions.
While a similar study to that of Ref. \cite{BH_Cong} could be undertaken following this approach with a discrete spectrum, here we will follow a different path. We will seek for solutions to Eq. \eqref{Qmass} for all real values of the mass $m_c$. This can only be possible if we search for them in a set much larger than the kinematic Hilbert space of the $b$-sector. The main idea is to take a domain of definition for our Hamiltonian constraint operator sufficiently small as to allow for a considerably large algebraic dual, and then impose the constraint by its dual action. A motivation for this is that, ideally, if one manages to construct solutions to the constraint for all values of the black hole mass, the resulting physical states would favor a continuous classical limit, at least for large masses, and one would not expect to encounter obstacles for transitions between very massive black holes, e.g. in evaporation processes.

In this spirit, let us now consider states $|\psi^{\tilde{\epsilon}_b}_{\delta_b}\rangle$ in the algebraic dual (in spite of the ket notation) of the dense set defined by the linear span of the eigenstates of $\hat{\tilde{p}}_b$ in our superselection sector. This allows for solutions that would not have found a place in the previous spectral analysis. Expressing those states in terms of the basis elements $|\tilde{\mu}_b \rangle$, we conclude that, to solve the constraint, their corresponding coefficients $\psi^{\tilde{\epsilon}_b}_{\delta_b}(\tilde{\mu}_b)$ must fulfill the recurrence relation 
\begin{eqnarray}
\label{eq: IV_rr}
\psi^{\tilde{\epsilon}_b}_{\delta_b}(\tilde{\epsilon}_b + 2n) &=& \mathfrak{A}[\tilde{\epsilon}_b + 2(n-1)]\psi^{\tilde{\epsilon}_b}_{\delta_b}[\tilde{\epsilon}_b + 2(n-4)]+\mathfrak{B}_{m_c}[\tilde{\epsilon}_b + 2(n-1)]\psi^{\tilde{\epsilon}_b}_{\delta_b}[\tilde{\epsilon}_b + 2(n-3)] \nonumber \\
&+&\mathfrak{C}_{\delta_b}[\tilde{\epsilon}_b + 2(n-1)]\psi^{\tilde{\epsilon}_b}_{\delta_b}[\tilde{\epsilon}_b+ 2(n-2)]+\mathfrak{D}_{m_c}[\tilde{\epsilon}_b + 2(n-1)]\psi^{\tilde{\epsilon}_b}_{\delta_b}[\tilde{\epsilon}_b + 2(n-1)],
\end{eqnarray}
where 
\begin{eqnarray}
\mathfrak{A}(\tilde{\mu}) &=& - \frac{f_{-}(\tilde{\mu} - 2)}{f_{-}(\tilde{\mu})}\frac{f_{-}(\tilde{\mu} - 4)}{f_{+}(\tilde{\mu})},\qquad \mathfrak{B}_{m_c}(\tilde{\mu}) = - \frac{f_{-}(\tilde{\mu} - 2)}{f_{-}(\tilde{\mu})}\frac{16 m_c}{ f_{+}(\tilde{\mu})}i,\\
\label{Drara}
\mathfrak{C}_{\delta_b}(\tilde{\mu}) &=&   \frac{1}{f_{-}(\tilde\mu)f_{+}(\tilde{\mu})}\left[16(\tilde{\mu}-2)^2  \delta_b^2 + f^2_{-}(\tilde{\mu} - 2) + f^2_{-}(\tilde{\mu})\right], \qquad  \mathfrak{D}_{m_c}(\tilde{\mu}) = \frac{16  m_c}{ f_{+}(\tilde{\mu})}i .
\end{eqnarray}
The above relation implies that all the coefficients are determined by four pieces of initial data. Nevertheless, Eq. \eqref{eq: IV_rr} for $n=3$ and $n=4$ provides two constraints between those four values, reducing the number of independent initial data to two. Using these constraints between the initial values, we obtain the following closed expression for the coefficients of the solution:
\begin{eqnarray}
\label{eq: IV_ac}
\psi^{\tilde{\epsilon}_b}_{\delta_b}(\tilde{\epsilon}_b + 2n) &=& [\mathfrak{A}(\tilde{\epsilon}_b + 6)\mathcal{T}_{\tilde{\epsilon}_b}(4,n) + \mathfrak{B}_{m_c}(\tilde{\epsilon}_b + 4) \mathcal{T}_{\tilde{\epsilon}_b}(3,n) + \mathfrak{C}_{\delta_b}(\tilde{\epsilon}_b + 2)\mathcal{T}_{\tilde{\epsilon}_b}(2,n)] \psi^{\tilde{\epsilon}_b}_{\delta_b}(\tilde{\epsilon}_b)\nonumber\\  
&+& \mathcal{T}_{\tilde{\epsilon}_b}(1,n)\psi^{\tilde{\epsilon}_b}_{\delta_b}(\tilde{\epsilon}_b +2), 
\end{eqnarray}
with
\begin{gather}
\mathcal{T}_{\tilde{\epsilon}_b}(u,v) = \sum_{\tilde{O}(u \rightarrow v)} \prod_{\{ q_u \}}\mathfrak{D}_{m_c}(\tilde{\epsilon}_b + 2q_u) \prod_{\{ r_u \}}\mathfrak{C}_{\delta_b}[\tilde{\epsilon}_b +2(r_u + 1)]
\prod_{\{ s_u \}}\mathfrak{B}_{m_c}[\tilde{\epsilon}_b + 2(s_u + 2)] \prod_{\{ t_u \}}\mathfrak{A}[\tilde{\epsilon}_b + 2(t_u + 3)].
\end{gather}
Here, $\tilde{O}(u \rightarrow v)$ is the collection of all possible paths connecting $u$ to $v$ by integer steps of no more than four units. For each element in this collection, the sets $\{q_u\}$, $\{r_u\}$, $\{s_u\}$, and $\{t_u\}$ denote the respective subsets of integers followed by a jump of one, two, three, or four units. 

The constructed solutions do not admit a smooth limit for large $\tilde{\mu}_b$ for the same reason explained before in the case of the operator $\hat{\Omega}_j$, namely, they split into two functions with support on different semilattices that display a rapidly varying relative phase. Expressing our state as $|\psi^{\tilde{\epsilon}_b}_{\delta_b}\rangle = |\psi^{\tilde{\epsilon}_b}_{\delta_b,1}\rangle \oplus i |\psi^{\tilde{\epsilon}_b}_{\delta_b,2}\rangle$, where $|\psi^{\tilde{\epsilon}_b}_{\delta_b,1}\rangle$ has support on $^{(4)}\mathcal{L}^{+}_{\tilde{\epsilon}_{b}}$ and $|\psi^{\tilde{\epsilon}_b}_{\delta_b,2}\rangle$ on $^{(4)}\mathcal{L}^{+}_{\tilde{\epsilon}_{b} + 2}$, we can analyze the limit separately for each of these two semilattices. Substituting this decomposition into Eq. \ref{Qmass} and solving carefully the resulting system for $|\psi^{\tilde{\epsilon}_b}_{\delta_b,1}\rangle$, we obtain
\begin{gather}
\label{eq: IV_phi1}
\left[\delta_b^2\hat{\Omega}_b^2 + \delta_b^4\hat{\tilde{p}}_b^2 + \delta_b^2\hat{\Omega}_b(\hat{\tilde{p}}_b)^{-2}\hat{\Omega}_b\hat{\tilde{p}}_b^2 + \hat{\Omega}_b(\hat{\tilde{p}}_b)^{-2}\hat{\Omega}_b^3 - 4m_c^2\hat{\Omega}_b(\hat{\tilde{p}}_b)^{-2}\hat{\Omega}_b\right] |\psi^{\tilde{\epsilon}_b}_{\delta_b,1}\rangle = 0.
\end{gather}
To arrive to this equation, we have only inverted the operator $\hat{\tilde{p}}_b^2$, something that is always possible in our semilattices. In particular, the above equation implies that $|\psi^{\tilde{\epsilon}_b}_{\delta_b,1}\rangle$ has real coefficients in the basis given by $|\tilde{\mu}_b\rangle$ (up to a constant, global phase). Although we will focus our discussion on $|\psi^{\tilde{\epsilon}_b}_{\delta_b,1}\rangle$, it should be noted that, for all $\delta_b\neq 0$, the other part of our state is completely fixed via the following relation: 
\begin{equation}
|\psi^{\tilde{\epsilon}_b}_{\delta_b,2}\rangle =- \frac{i}{2m_c\delta_b^2}(\hat{\tilde{p}}_b)^{-2}\hat{\Omega}_b \left[\hat\Omega^2_b+\delta_b^{2}\hat{\tilde{p}}_b^2-4m_c^2\right]|\psi^{\tilde{\epsilon}_b}_{\delta_b,1}\rangle .
\end{equation}

Work experience leads us to propose a smooth limit of the form 
\begin{equation}\label{apm}
\underline{\psi}^{\tilde{\epsilon}_b}_{\delta_b,1(\pm)}(\tilde{\mu}_b)=\chi(\tilde{\mu}_b) \exp[a_{(\pm)}\tilde{\mu}_b], \qquad a_{(\pm)}=\frac{1}{2}\\\ln\left(\sqrt{1+\delta_b^2}\pm |\delta_b| \right).
\end{equation}  
The exponents $a_{(\pm)}$ are the two real roots that remove the otherwise dominant, quadratic contribution in $\tilde{\mu}_b$ for large values of this quantity in the WDW counterpart of Eq. \eqref{eq: IV_phi1}. Note that $a_{(+)}$ is positive for all values of $\delta_b$ and $a_{(-)}=-a_{(+)}$. The procedure for obtaining the WDW equation is explained in Appendix \ref{sec: App_WDW}. Substituting the ansatz $\chi(\tilde{\mu}_b) = \tilde{\mu}_b^{d}$ (up to subdominant terms) in the equation resulting for $\chi$ in this WDW limit, we conclude that, for any nonzero value of $\delta_b$, the exponent $d$ can be either of the following complex conjugate constants \footnote{We can ignore from all considerations the value $\delta_b=0$, e.g. by demanding that the function $\tilde{K}_b$, with which $\delta_b$ coincides on physical solutions, be nowhere zero. In any case, $\delta_b=0$ is a point of zero Lebesgue measure.}:
\begin{equation}
\label{eq: IV_d}
 d_s = -1 + i \; s \; \sqrt{\frac{m_c^2}{1 + \delta_b^2}}, \qquad s=1 \quad \text{or} \quad -1. 
\end{equation} 

Taking into account that the solutions to Eq. \eqref{eq: IV_phi1} are real, modulo a constant phase, we conclude that, out of the four complex WDW behaviors that we have presented [given by the two possible values of the labels $(\pm)$ and $s$], only real linear combinations obtained with complex conjugate pairs of powers $d_s$ are admissible: 
\begin{eqnarray}\label{asym}
&&\left[\zeta_{(+)}\chi_{1}(\tilde{\mu}_b) + \zeta_{(+)}^{\ast} \chi_{-1}(\tilde{\mu}_b)\right] \exp[a_{(+)}\tilde{\mu}_b],\\
&&\left[\zeta_{(-)}\chi_{1}(\tilde{\mu}_b) + \zeta_{(-)}^{\ast} \chi_{-1}(\tilde{\mu}_b)\right] \exp[a_{(-)}\tilde{\mu}_b],
\end{eqnarray}
where $\zeta_{(\pm)}$ are complex constants and $\chi_{s}(\tilde{\mu}_b)=\tilde{\mu}_b^{d_s}$. In principle, any real linear combination of these two contributions could be valid, allowing for two real independent pieces of data up to normalization. Nonetheless, we note that the second contribution is subdominant for large $\tilde{\mu}_b$ with respect to the first one (in the considered superselection sector with positive orientation), and furthermore exponentially damped, because we have seen that $a_{(+)}>0>a_{(-)}$. Consequently, except for a critical case, all solutions display an asymptotic limit of the form \eqref{asym}. 

We have seen that the construction of our solutions on the $b$-sector depends on two initial data. The above analysis indicates that any attempt to restrict the choice of these data by imposing a specific asymptotic behavior would not be feasible, because this behavior is essentially unique (except for a case with exponentially subdominant behavior with respect to any other solution, no matter how small the amplitude of the latter in a linear superposition). 
However, it is possible to reduce the number of independent data to only one by the following method. If we consider the direct classical counterpart of the constraint \eqref{Qmass}, we obtain two possible roots,
\begin{equation}
\Omega_b^{\pm}(\tilde{p}_b)=- m_c \pm \sqrt{m_c^2 - \delta_b^2\tilde{p}_b^2}.
\end{equation} 
In the limit in which $\delta_b$ is asymptotically small (keeping $\tilde{p}_b$ fixed), $\Omega_b^{-}$ is equal to $-2m_c$ at dominant order, while $\Omega_b^{+}$ becomes negligible. Taking into account that, up to a constant multiplicative factor, $\Omega_b$ equals the partial Hamiltonian $O_b$ at dominant order, and therefore the black hole mass on-shell, we see that $\Omega_b^{-}$ is the physically preferred solution to the constraint equation. We can try and impose a similar relation between our two pieces of initial data for the coefficients of $| \psi^{\tilde{\epsilon}_b}_{\delta_b} \rangle$. Namely, we can demand that $\langle \tilde{\epsilon}_b | \hat{\Omega}_b -\Omega_b^{-}(\hat{\tilde{p}}_b) | \psi^{\tilde{\epsilon}_b}_{\delta_b} \rangle = 0$ on our solutions. In this way, we obtain a condition on the first two coefficients of the state in the superselection sector for $b$ characterized by $\tilde{\epsilon}_b $.  This condition can be written as 
\begin{equation}
\psi^{\tilde{\epsilon}_b}_{\delta_b}(\tilde{\epsilon}_b + 2) = \frac{1}{2}\mathfrak{D}_{m_c}(\tilde{\epsilon}_b)\left[1+\sqrt{1-\left(\frac{\delta_b\gamma \tilde{\epsilon}_b}{2 m_c}\right)^2}\right] \psi^{\tilde{\epsilon}_b}_{\delta_b}(\tilde{\epsilon}_b),
\end{equation}
where $\mathfrak{D}_{m_c}$ is given by Eq. \eqref{Drara}. When we substitute this relation into Eq. \eqref{eq: IV_ac}, our algorithm for the construction of solutions gives 
\begin{eqnarray}
\frac{\psi^{\tilde{\epsilon}_b}_{\delta_b}(\tilde{\epsilon}_b+2n)}{\psi^{\tilde{\epsilon}_b}_{\delta_b}(\tilde{\epsilon}_b) } &=&   \frac{1}{2} \left[1+\sqrt{1-\left(\frac{\delta_b\gamma \tilde{\epsilon}_b}{2 m_c}\right)^2}\,\right] \mathfrak{D}_{m_c}(\tilde\epsilon_b)\mathcal{T}_{\tilde{\epsilon}_b}(1,n) + \mathfrak{C}_{\delta_b}(\tilde{\epsilon}_b + 2)\mathcal{T}_{\tilde{\epsilon}_b}(2,n) + \mathfrak{B}_{m_c}(\tilde{\epsilon}_b + 4)\mathcal{T}_{\tilde{\epsilon}_b}(3,n)    \nonumber\\ 
&+&   \mathfrak{A}(\tilde{\epsilon}_b + 6)\mathcal{T}_{\tilde{\epsilon}_b}(4,n).
\end{eqnarray}
Thus, the coefficients of $|\psi^{\tilde{\epsilon}_b}_{\delta_b}\rangle$ are determined by a single initial value. 

\section{Physical states\label{sec: V}}

To construct physical states, we will follow the approach developed in Ref. \cite{BH_GAB}, adapted to our specific case. Recall that we have searched for solutions to the geometric part of the Hamiltonian constraint in the algebraic dual of the eigenstates of $\hat{\tilde{p}}_j$, for each generalized eigenspace of the operators $\hat{\delta}_j$. We can therefore express the resulting states $|\xi_p\rangle$ in terms of the basis elements $|\tilde{\mu}_b,\tilde{\mu}_c,\delta_b,\delta_c\rangle$ (adopting again a ket notation). Together with our results, that determine the solution $\psi^{\tilde{\epsilon}_b}_{\delta_b}(\tilde{\mu}_b)$ in the $b$-sector for each value of $\delta_b$ and $m_c$, we then obtain
\begin{gather}\label{phystate}
|\xi_p\rangle=\int_\mathbb{R}\text{d}\delta_b\int_\mathbb{R}\text{d}\delta_c\int_\mathbb{R}\text{d}m\sum_{\tilde{\mu}_b,\tilde{\mu}_c} \xi_p(\delta_b,\delta_c,m) \psi^{\tilde{\epsilon}_b}_{\delta_b}(\tilde{\mu}_b)e^{\tilde{\epsilon}_c}_{mL_o}(\tilde{\mu}_c) \,| \tilde\mu_b,\tilde\mu_c,\delta_b,\delta_c\rangle.
\end{gather}
The sum over $\tilde{\mu}_j$ is taken over all points in the semilattice of our superselection sector for the geometry, determined by $\tilde{\epsilon}_j$, and we have changed integration variables from $m_c$ to $m=m_c/L_o$. The integral over $m$ runs over the whole real axis because this is the continuous interval of definition for $m_c$ (the generalized eigenvalue of $\hat{\Omega}_c$) and because we have succeeded in finding solutions to the Hamiltonian constraint for all values of this mass by permitting a sufficiently broad habitat for them. The integrals over the $\delta$-parameters correspond to the expansion of the quantum state in the basis with elements that are generalized eigenstates of the operators $\hat{\delta}_j$. In general, states of the form \eqref{phystate} do not yet satisfy the constraints on $\delta_j$ introduced when extending the phase space (they only satisfy the quantum Hamiltonian constraint).

The generalized function $\xi_p$ in the above expansion totally determines the state. Ultimately, in order to satisfy the commented constraints on $\delta_j$ in our extended phase space formalism, it must have the specific form
\begin{gather}\label{wavef}
\xi_p(\delta_b,\delta_c,m) = \xi(m)\;\delta[\delta_b - \tilde{K}_b(m)]\delta[\delta_c - \tilde{K}_c(m)].
\end{gather}
Apart from the product of Dirac deltas that identify the regularization parameters $\delta_j$ with the functions $\tilde{K}_j(m)$ introduced in Eq. \eqref{eq: II_Ktilde}, we see that the state is characterized by the wave function of the black hole mass $\xi(m)$. So, after having imposed all the constraints, the final expression of any state is 
\begin{gather}\label{phystatekk}
|\xi_p\rangle=\int_\mathbb{R}\text{d}m\sum_{\tilde{\mu}_b,\tilde{\mu}_c}\xi(m)\psi^{\tilde{\epsilon}_b}_{\delta_b}(\tilde{\mu}_b)|_{\delta_b \!=\! \tilde{K}_b(m)}e^{\tilde{\epsilon}_c}_{mL_o}(\tilde{\mu}_c) \,| \tilde\mu_b,\tilde\mu_c,\delta_b = \tilde{K}_b(m),\delta_c \!=\! \tilde{K}_c(m)\rangle.
\end{gather}
We recall that the limit of large black hole masses corresponds to negligibly small values of the $\delta$-parameters. We can construct states in this context by considering wave functions $\xi(m)$ that are peaked on the region of large masses. Those profiles exist, because $m$ is a continuous variable in that region.

Finally, let us discuss how we can endow our set of physical states with a Hilbert space structure. From the form of these states in Eqs. \eqref{phystate} and \eqref{wavef}, it is clear that, for each of the $\delta$-parameters, we have to eliminate two redundant (squared) Dirac deltas from the kinematic inner product. This leads us to consider instead the norm in the geometric part of the kinematic Hilbert space. According to Eq. \eqref{phystatekk}, and using that the eigenstates of $\hat{\Omega}_c$ are normalized to the Dirac delta in $m_c=m L_o$, this gives
\begin{equation}
\langle \varsigma_p |\xi_p\rangle_{\text{geom}} = \frac{1}{L_o} \int_\mathbb{R} \text{d}m \; \varsigma^{\ast}(m)\xi(m) \sum_{\tilde{\mu}_b}  \left|\psi^{\tilde{\epsilon}_b}_{\delta_b}(\tilde{\mu}_b)|_{\delta_b \!=\! \tilde{K}_b(m)}\right|^2 ,
\end{equation}
where $\varsigma(m)$ is the mass wave function of the physical state $|\varsigma_p\rangle$. Our asymptotic analysis in Sec. \ref{sec: IV} [and, in particular, Eqs. \eqref{apm} and \eqref{eq: IV_d}] proves that the squared complex norm of $\psi^{\tilde{\epsilon}_b}_{\delta_b}$ generally grows for large $\tilde{\mu}_b$ as 
\begin{equation} \label{diver}
 {\tilde{\mu}_b}^{-2}\exp[\tilde{\mu}_b \ln(\sqrt{1+\delta_b^2}+|\delta_b|)].
\end{equation}
The sum over $\tilde{\mu}_b$ would then diverge. We can remove this divergence in different manners. The simplest way is to absorb the whole divergent sum by a redefinition of $\xi(m)$, taking then as physical inner product for the redefined mass wave functions that of square integrable functions over the real line. A less obvious possibility is to modify the kinematic inner product in the $b$-sector. We can change the normalization of the basis of eigenstates $|\tilde{\mu}_b\rangle$ from a Kronecker delta to its product by the inverse of the exponential factor in Eq. \eqref{diver}, making the considered sum convergent. However, the required factor depends on $\delta_b$, which equals $\tilde{K}_b(m)$ on physical states. Hence, the suggested change of inner product in the $b$-sector would depend on the mass. An interesting situation is found when $|\tilde{K}_b|$ is a function bounded from above (which is the case for large $m$ in the AOS model). Let us call this bound $B$. Then, for all values of the black hole mass, the exponential term in Eq. \eqref{diver} grows always less rapidly than $(\sqrt{1+ B^2}+ B)^{{\tilde{\mu}_b}}$. So, we can introduce a physical inner product by defining   
\begin{equation}
\langle \varsigma_p |\xi_p\rangle_{\text{phys}} = \frac{1}{L_o} \int_\mathbb{R} \text{d}m \; \varsigma^{\ast}(m) \xi(m) \sum_{\tilde{\mu}_b}  (\sqrt{1+B^2}+ B)^{-{\tilde{\mu}_b}} 
\left|\psi^{\tilde{\epsilon}_b}_{\delta_b}(\tilde{\mu}_b)|_{\delta_b \!=\! \tilde{K}_b(m)}\right|^2 .
\end{equation} 

The solutions constructed at the beginning of this section form a Hilbert space with the discussed inner product. Linear operators in this Hilbert space provide the physical observables. Clearly, one such observable is the black hole mass.

\section{Conclusions\label{sec: VI}}

We have presented a complete quantization of the extended phase space formulation of the AOS model that provides a description of the interior geometry of a nonrotating black hole in LQC. In particular, we have represented the constraints of the system as quantum operators. Two of these constraints impose conditions on the regularization parameters of the model and the other one accounts for the Hamiltonian constraint on the geometry. For this last constraint, and in contrast with previous studies \cite{BH_GAB}, we have adopted a densitization similar to the most standard one in LQG. In this way we have avoided handling with inverse operators that would not admit a straightforward definition in the usual basis of triad eigenstates for LQC. With this representation, we have discussed the existence of solutions to all of the constraints. The quest for solutions to the Hamiltonian constraint has led us to consider a habitat for them much broader than the kinematic Hilbert space of the geometry. We have focused the attention on the linear span of the eigenstates of some suitably rescaled triad variables and searched for solutions in its algebraic dual. Thanks to this, we have been able to construct solutions for all possible values of the black hole mass. This mass is a Dirac observable of the quantum system. The availability of solutions for a continuous range of masses favors the validity of a semiclassical regime of large black hole masses and facilitates the consideration of the limit of negligibly small regularization parameters, as this occurs in the AOS model when the mass tends to infinity. This limit can be approached in our quantum model with no intrinsic discreteness. The consideration and determination of solutions allowing for black hole masses in the whole real line is a distinctive feature of our quantization in comparison to the noteworthy proposal of Refs. \cite{BH_Con,BH_Cong}, for which this mass presents a discrete spectrum. A real interval of masses dissipates the expectation of black hole remnants at the end of evaporation processes, justified by the existence of a minimum nonzero mass when this quantity has a discrete nature. 

For all real masses, we have analyzed the asymptotics of our solutions to the Hamiltonian constraint in the WDW limit of large (rescaled) triad variables. We have determined their divergent behavior with respect to the kinematic inner product, and suggested some possible ways to absorb it, at least in some
cases. Based on this asymptotic analysis, we have discussed how we can introduce an inner product in our set of solutions, obtaining in this way a physical Hilbert space. Among the possible operators that can be defined in this space, which are the observables of the quantum theory, the black hole mass is certainly a notable one. 
 
An important property of our quantization is that our kinematic representation of the extended phase space, after a convenient rescaling of the triad variables, decouples the geometric Hilbert space from that of the regularization parameters. More concretely, the total kinematic Hilbert space for the representation of the constraints is a tensor product of spaces, in which the geometric one is always the same, regardless of the values of the regularization parameters in the quantum system. This fact extremely simplifies the spectral analysis of the geometric operators that are needed in the quantum constraints (e.g. compared to Ref. \cite{BH_Cong}), because their domains of definition and basic properties do not change from one to another of the generalized eigenspaces of those parameters. Moreover, our algorithm for the construction of solutions to the Hamiltonian constraint depends very mildly on those parameters, and always by means of analytic functions of them.

For any real value of the black hole mass, this algorithm fixes the whole solution in terms of some initial data. The dependence on the $c$-sector is totally specified by the mass of the black hole (up to normalization). With respect to the $b$-dependence, on the other hand, we have first determined the solution using two initial data, which have been chosen as the values at the two points that are closer to the origin in the semilattice of the considered superselection sector. We have then reduced the number of free data to only one, that can be given by normalization. This reduction has been possible by requiring a relation between the two otherwise independent data that parallels the classical relation that has a correct physical behavior for small regularization parameters. This procedure determines the solutions (up to irrelevant global factors) once the black hole mass is known. The most general solution, which is simply a superposition formed with different values of this black hole mass, can then be characterized by its mass profile, i.e., by the mass wave function.      

In a future research, it would be interesting to include matter in the model and discuss its use as an internal clock. An even more appealing possibility is to introduce metric and/or matter perturbations in these quantum black holes, exploring if the system allows for transitions between black hole states and studying the evolution of the perturbations. This perturbed system could be treated using the hybrid formalism for LQC \cite{hyb-review}. The ultimate idea would be to apply this kind of analysis to Hawking radiation (see e.g. Ref. \cite{GP}) or to quasinormal modes \cite{quasi}, investigating the role that a continuous interval of black hole masses plays in the theory. Quasinormal modes describe the gravitational waves emitted during ringdown \cite{ringdown} in the last stages of black hole mergers. Our work could open a door to the study and estimation of LQG modifications to their gravitational emission.

\acknowledgments

This work was partially supported by Project No. MICINN PID2020-118159GB-C41 from Spain, Grants NSF-PHY-1903799, NSF-PHY-2206557, and funds of the Hearne Institute for Theoretical Physics. The authors are grateful to A. Garc\'{\i}a-Quismondo, D. Mart\'{\i}n de Blas, and J. Olmedo for enlightening discussions. They also want to thank I. Agullo and D. Kranas for feedback, and R. Gambini and C. Sopuerta for helpful conversations.

\appendix

\section{Spectral analysis of the operators\label{sec: App_SA}}

Let us first demonstrate that the operator $\hat{\Omega}^2_j$ is essentially self-adjoint. We choose its domain of definition as 
\begin{equation}
\mathcal{D}_j=D(\hat{\Omega}^2_j) = \oplus_{\tilde{\varepsilon}_j}(^{(4)}\text{Cyl}_{\tilde{\varepsilon}_j}^{+} \cup\! ^{(4)}\text{Cyl}_{4-\tilde{\varepsilon}_j}^{-}),\qquad \,^{(4)}\text{Cyl}_{\tilde{\varepsilon}_{j}}^{\pm} = \text{span} \{ |\tilde{\mu}_{j}\rangle : \tilde{\mu}_{j} \in\! ^{(4)}\mathcal{L}_{\tilde{\varepsilon}_{j}}^{\pm} \}.
\end{equation}
Its action on the Hilbert space is decomposed into two symmetric operators, $\hat{H}^1_j = \hat{\mathcal{N}}_{4\delta_j}\hat{C}^{4}_j + \hat{\mathcal{N}}_{-4\delta_j}\hat{C}^{-4}_j$ and $\hat{C}^{0}_j$, with 
\begin{equation}
C^{\pm4}_j(\tilde{\mu}_j)=-\omega_j^2 f_{\pm}(\tilde{\mu}_j) f_{\pm}(\tilde{\mu}_j\pm 2),\qquad C^{0}_j(\tilde{\mu}_j)=\omega_j^2 [ f_{+}^2(\tilde{\mu}_j) + f_{-}^2(\tilde{\mu}_j)],
\end{equation}
and their corresponding operators defined as functions of the multiplicative operator $\hat{\tilde{\mu}}_j$, proportional to $\hat{\tilde{p}}_j$. We recall that $f_{\pm}(\tilde{\mu}_j) $ is given in Eq. \eqref{fpm} and $\omega_b = 1/8$, $\omega_c=1/4$. It is straightforward to see that $\hat{\tilde{\mu}}_j$ and, therefore, $\hat{C}^{0}_j$, with domain $\mathcal{D}_j$, are essentially self-adjoint operators. Based on this, if the inequality $\|\hat{H}^{1}_j \Psi\|^2 \leq \|\hat{C}^{0}_j \Psi\|^2 + \alpha^2_j \|\Psi\|^2$ holds for certain positive constant $\alpha^2_j$ and any $\Psi \in \mathcal{D}_j$, then the Kato-Rellich theorem \cite{Kato} ensures that $\hat{\Omega}^2_j$ is essentially self-adjoint. Using the triangle inequality, we get $\|\hat{H}^{1}_j \Psi\|^2 \leq \| \hat{\mathcal{N}}_{+4}\hat{C}^{4}_j\Psi\|^2 + \|\hat{\mathcal{N}}_{-4}\hat{C}^{-4}_j\Psi \|^2$. On the other hand, the functions $C^{\pm 4}_j(\tilde{\mu}_j)$ and $C^0_j(\tilde{\mu}_j)$, for all the allowed values of $\tilde\mu_j$ in $\mathbb{R}_{\pm}$, satisfy the inequality 
\begin{align}
&\left[ C_j^{4}(\tilde{\mu}_j)\right]^2 + \left[ C_j^{-4}(\tilde{\mu}_j)\right]^2 \leq \left[ C_j^{0}(\tilde{\mu}_j)\right]^2- 4\omega_j^4z_{\pm}(\tilde{\mu}_j),\\
\label{eq: App_f}
&z_{\pm}(\tilde{\mu}_j) = \frac{1}{2}f_+^2(\tilde{\mu}_j)f_-^2(\tilde{\mu}_j) +(|\tilde{\mu}_j||\tilde{\mu}_j\pm 2|-|\tilde{\mu}_j\pm 2||\tilde{\mu}_j\pm 4|)f_{\pm}^2(\tilde{\mu}_j) .
\end{align}
We then arrive at an inequality of the desired type for $\|\hat{H}^{1}_j \Psi\|^2$ by taking $\alpha^2_j = 4\omega_j^4|z_{\text{min}}|$, where $z_{\text{min}}$ denotes the coincident minimum value of the functions $z_{\pm}(\tilde{\mu}_j)$, which can be shown to be $z_{\text{min}}\approx -887$. Our proof of self-adjointness has followed in part the analysis carried out in Refs. \cite{K&L_I, K&L_II} in the context of LQC.  
 
Actually, it can be shown that $\hat{\Omega}^2_j$ is essentially self-adjoint in each domain $^{(4)}\text{Cyl}_{\tilde{\varepsilon}_j}^{\pm}$. This can be proven by contradiction. Consider the deficiency index equation $(\hat{\Omega}_j^{2\dagger}-\rho)|\Phi\rangle = 0$ in the considered domain, where $\rho$ is an arbitrary complex number with nonvanishing imaginary part. We recall that the dimension of the subspace of solutions to this equation is constant in each half-plane of the complex numbers $\rho$ with the same sign of the imaginary part. Now, any nontrivial solution immediately provides a solution to the deficiency index equation in the total domain, by letting it vanish outside the original semilattice. However, this cannot be true because, with this total domain, the operator is known to be essentially self-adjoint, so that its deficiency indices vanish. Therefore the only solution in each of the restricted domains $^{(4)}\text{Cyl}_{\tilde{\varepsilon}_j}^{\pm}$ must be the trivial one, ensuring self-adjointness also in these cases.

The great similarity between $\hat{\Omega}_j^2$, with domain $\mathcal{D}_j$, and the geometric part of the Hamiltonian constraint in homogeneous and isotropic cosmologies \cite{MMO} leads us to anticipate that their essential and absolutely continuous spectra coincide and are equal to $\left[0,\infty\right)$.
This can be actually proven by combining the results of Refs. \cite{MMO,Prescrip_G}. Specifically, if we identify $\tilde{\mu}_j$ with the values of the volume variable $v$ in LQC, the action of $\hat{\Omega}^2_j$ in our model coincides with that of the operator defined in flat homogeneous and isotropic cosmological spacetimes using the so called simplified MMO prescription (sMMO) \cite{Prescrip_G}, up to a multiplicative factor. This cosmological operator differs from a particular case analyzed in Ref. \cite{K&L_I} (the so-called sLQC prescription \cite{K&P}) by a compact perturbation only supported around $v=0$. Since that case presents an absolutely continuous and positive definite spectrum, equal to the essential one, in
$^{(4)}\text{Cyl}_{\tilde{\varepsilon}_j}^{+} \cup\! ^{(4)}\text{Cyl}_{4-\tilde{\varepsilon}_j}^{-}$, the same holds for $\hat{\Omega}^2_j$.
Finally, following the same arguments as in LQC \cite{MMO}, we also arrive at these spectral properties for the restriction of $\hat{\Omega}_j^2$ to the domain $^{(4)}\text{Cyl}_{\tilde{\varepsilon}_j}^{\pm}$.

Let us now demonstrate that the operator $\hat{\Omega}_j$ is also essentially self-adjoint in $\mathcal{D}_j$. A proof by contradiction shows that it is so in the domain $\!^{(2)}\text{Cyl}_{\tilde{\epsilon}_j}^{\pm} = \!^{(4)}\text{Cyl}_{\tilde{\epsilon}_j}^{\pm} \cup \!^{(4)}\text{Cyl}_{\tilde{\epsilon}_j +2}^{\pm}$. It suffices to consider the deficiency index equation for this domain, 
$(\hat{\Omega}^{\dagger}_j-\sigma)|\Phi\rangle = 0$, where $\sigma$ is an arbitrary complex number with nonvanishing imaginary part.
If a nontrivial solution exists, this would imply that $(\hat{\Omega}_j^{2\dagger} - \sigma^2)|\Phi\rangle = 0$ for any such $\sigma$. On the other hand, we have proven that the square operator is essentially self-adjoint in any of the domains $^{(4)}\text{Cyl}_{\tilde{\varepsilon}_j}^{\pm}$.
Then, there  is no way to satisfy the equality, and we conclude that $\hat{\Omega}_j$ has to be essentially self-adjoint in $\!^{(2)}\text{Cyl}_{\tilde{\epsilon}_j}^{\pm}$.
With similar arguments, it can be shown that the operator is also essentially self-adjoint in the domain $\mathcal{D}_j$. Indeed, taking the direct sum of the partial domains $^{(2)}\text{Cyl}_{\tilde{\epsilon}_j}^{\pm}$, self-adjointness must hold, otherwise there would exist a nontrivial solution for the deficiency index equation of $\hat{\Omega}_j$ in at least one of the partial domains, leading to a contradiction.

The similarity between $\hat{\Omega}_j$ and the geometric operators that appear in the Hamiltonian constraint of Bianchi I spacetimes in LQC \cite{BI_MGT} indicates that the spectrum of $\hat{\Omega}_j$, with domain $^{(2)}\text{Cyl}_{\tilde{\epsilon}_j}^{\pm}$, is the entire real line. This conclusion is supported by the results about the operator $\hat{\beta}_{\lambda}$ studied in Ref. \cite{BH_Cong}, which plays there the same role as $\hat{\Omega}_j$ here.

Finally, let us prove that the operator $\hat{\mathcal{Q}}_b(a)$ introduced in the left-hand side of Eq. \eqref{Qmass} has a discrete spectrum. Direct inspection shows that $\langle \hat{\mathcal{Q}}_b(a) \rangle_{\psi} \geq \delta_b^2 \langle \hat{\tilde{p}}_b^2 \rangle_{\psi}$ for all states $|\psi\rangle$ in its domain, dense in the kinematic Hilbert space of the $b$-sector. Therefore, the operator has a lower bound independent of the value of $a$ as long as $\delta_b$ does not vanish. According to the min-max principle (see e.g. Ref. \cite{G&P}), the pure discreteness of the spectrum of $\hat{\mathcal{Q}}_b(a)$ is then guaranteed for any value of $a$ except when $\delta_b$ becomes equal to zero.

Moreover, each of its eigenvalues $\rho$ describes an analytic curve as a function of $a$. This is a consequence of the analyticity of the operator in $a$, in the sense of Kato \cite{Kato}. Our demonstration is based on the proof of analyticity of the operator $\hat{\mathfrak{h}}^{(m)}$ defined in Ref. \cite{BH_Cong}. Owing to the similarities between these two operators, we will only outline the most relevant steps of the proof. Consider the sesquilinear form $T_a(\psi,\varphi) = \langle \psi | \hat{\mathcal{Q}}_b(a) | \varphi \rangle$, defined for all states $| \psi \rangle$ and $| \varphi \rangle$ of the kinematic Hilbert space for the $b$-sector belonging to the domain $\mathcal{D}_b$ of the operator $\hat{\mathcal{Q}}_b(a)$. One can easily see that the form $T_a$ is symmetric, bounded from below, and closable. We call its closure
$\overline{T_a}$ and it is the extension of $T_a$ to the closure of $\mathcal{D}_b$ with respect to the graph norm. Since the graph norm depends on $a$, in general, the closure of $\mathcal{D}_b$ could also depend on it. However, this is not the case, because the graph norm, for any value of $a$, is equivalent to the norm $\|\psi\|_{+}^2 = \langle \psi | \hat{\tilde{p}}_b^2 | \psi \rangle + \langle \psi | \psi \rangle$, and the latter does not depend on $a$. This result allows us to prove that $\overline{T_a}(\psi,\psi)$ is an analytic function of $a$ for any $|\psi\rangle \in \overline{\mathcal{D}_b}$, with the closure taken with respect to the norm $\|\cdot\|_{+}$. That this is so follows from the fact that 
\begin{equation}
T_a(\psi,\psi) = \langle \psi | \hat{\Omega}_b^2 + a^2 + 2a\hat{\Omega}_b + \delta_b^2\hat{\tilde{p}}_b^2 | \psi \rangle
\end{equation} 
is a linear combination of terms with analytic coefficients in $a$, where the values $\langle \hat{\tilde{p}}_b^2 \rangle_{\psi}$, $\langle \hat{\Omega}_b \rangle_{\psi}$, and $\langle \hat{\Omega}_b^2 \rangle_{\psi}$ are well defined for all $|\psi\rangle \in \overline{\mathcal{D}_b}$. Thus, we conclude that $\overline{T_a}$ is a holomorphic family of type $(a)$ in the sense of Kato \cite{Kato}. Additionally, since $\overline{T_a}$ is symmetric and closed, there exists an essentially self-adjoint operator $\overline{\hat{\mathcal{Q}}_b(a)}$ associated with it. In fact, this operator is the Friedrichs extension of $\hat{\mathcal{Q}}_b(a)$, and it forms a holomorphic family of type $(B)$ in the sense of Kato \cite{Kato}. The analyticity of its eigenvalues with respect to $a$ is a direct consequence of this fact \cite{R&S}.

\section{WDW limit \label{sec: App_WDW}}

In this appendix we will derive the counterpart of Eqs. \eqref{eq: III_O^2} and \eqref{eq: IV_phi1} in the WDW theory, understood as the limit for large values of $\tilde{\mu}_j$, in which the separation between points in the semilattices where the states have support become comparatively small. Our approach involves an asymptotic expansion, up to order $O(\tilde{\mu}_j^{-1})$, of the coefficients of the difference equations that determine the action of the studied operators.

From this perspective, the action of $\hat{\Omega}_j^2$ contains three contributions. The asymptotic expansion of their coefficients is given by 
\begin{equation}
f_{\pm}(\tilde{\mu}_j \pm 2)f_{\pm}(\tilde{\mu}_j) = 4\tilde{\mu}_j^2 \pm 16\tilde{\mu}_j + 8 + O(\tilde{\mu}_j^{-1}),\qquad 
f_{\pm}^2(\tilde{\mu}_j) = 4\tilde{\mu}_j^2 \pm 8 \tilde{\mu}_j .
\end{equation} In addition, the generalized eigenfunctions of the operator are expanded in the series 
\begin{equation}
e_{m_j^2}^{\tilde{\varepsilon}_j}(\tilde{\mu}_j \pm 4) = e_{m_j^2}^{\tilde{\varepsilon}_j}(\tilde{\mu}_j) \pm 4\partial_{\tilde{\mu}_j}e_{m_j^2}^{\tilde{\varepsilon}_j}(\tilde{\mu}_j) +8\partial_{\tilde{\mu}_j}^2e_{m_j^2}^{\tilde{\varepsilon}_j}(\tilde{\mu}_j) + \cdots.
\end{equation} 
By substituting the previous expansions into the eigenvalue equation of $\hat{\Omega}_j^2$ and neglecting all the contributions of order $O(\tilde{\mu}_j^{-1})$, with powers of $\tilde{\mu}_j\partial_{\tilde{\mu}_j}$ assumed to be of order one, we arrive at the following differential equation for the WDW limit $\underline{e}_{m_j^2}(\tilde{\mu}_j)$ of the eigenfunctions: 
\begin{equation}
m_j^2 \underline{e}_{m_j^2}(\tilde{\mu}_j)=- 16\omega_j^{2}[4\tilde{\mu}_j^2\partial^2_{\tilde{\mu}_j} + 8\tilde{\mu}_j\partial_{\tilde{\mu}_j} + 1] \underline{e}_{m_j^2}(\tilde{\mu}_j).
\end{equation} 
The expression used in Sec. \ref{ssec: WDW} follows from this one.

On the other hand, the action of the operator considered in Eq. \eqref{eq: IV_phi1} has five contributions. The asymptotic expansion of their coefficients is given by
\begin{eqnarray}
&&(\tilde{\mu}_b \pm 2)^{-2}f_{\pm}(\tilde{\mu}_b) f_{\pm}(\tilde{\mu}_b \pm 2)f_{\pm}(\tilde{\mu}_b \pm 4)f_{\pm}(\tilde{\mu}_b \pm 6)=16[\tilde{\mu}_b^2 \pm 12\tilde{\mu}_b + 32 ]+ O(\tilde{\mu}_b^{-1}),\\
&&(\tilde{\mu}_b \pm 2)^{-2} f_{\pm}(\tilde{\mu}_b)[f_{\pm}^3(\tilde{\mu}_b \pm 2) + f_{\pm}^2(\tilde{\mu}_b \pm 4)f_{\pm}(\tilde{\mu}_b \pm 2)+f_{\pm}^2(\tilde{\mu}_b)f_{\pm}(\tilde{\mu}_b \pm 2)]  \nonumber\\
&&  +(\tilde{\mu}_b \mp 2)^{-2}f_{\pm}(\tilde{\mu}_b\pm 2) f_{\pm}(\tilde{\mu}_b) f^{2}_{\mp}(\tilde{\mu}_b)= 64[\tilde{\mu}_b^2 \pm 6\tilde{\mu}_b + 10] + O(\tilde{\mu}_b^{-1}), 
\\
&& (\tilde{\mu}_b - 2)^{-2}[f_{-}^4(\tilde{\mu}_b)+ f_{-}^2(\tilde{\mu}_b)f_{+}^2(\tilde{\mu}_b) + f_{-}^2(\tilde{\mu}_b)f_{-}^2(\tilde{\mu}_b-2)] + (\tilde{\mu}_b + 2)^{-2}[f_{+}^4(\tilde{\mu}_b) + f_{+}^2(\tilde{\mu}_b)f_{-}^2(\tilde{\mu}_b) \nonumber\\ 
&&+ f_{+}^2(\tilde{\mu}_b)f_{+}^2(\tilde{\mu}_b+2)] = 32[3\tilde{\mu}_b^2 + 8 ]+ O(\tilde{\mu}_b^{-1}).
\end{eqnarray}
Introducing these asymptotic values into Eq. \eqref{eq: IV_phi1} we get a difference equation for the coefficients of $|\psi^{\tilde{\epsilon}_b}_{\delta_b,1}\rangle$ of the form
\begin{eqnarray}
&&16(2 \delta_b^2 - 8 m_c^2  + 1)\psi^{\tilde{\epsilon}_b}_{\delta_b,1}(\tilde{\mu}_b) - 8(8 \delta_b^2 - 8 m_c^2 + 5)\Delta_4^{+}(\tilde{\mu}_b) + 32\Delta_8^{+}(\tilde{\mu}_b)+12\tilde{\mu}_b[\Delta_8^{-}(\tilde{\mu}_b)-2(2 \delta_b^2 + 1)\Delta_4^{-}(\tilde{\mu}_b)] \nonumber \\
&&+ \tilde{\mu}_b^2[2(8 \delta_b^4 + 8 \delta_b^2 + 3)\psi^{\tilde{\epsilon}_b}_{\delta_b,1}(\tilde{\mu}_b)- 4(2 \delta_b^2 + 1)\Delta_4^{+}(\tilde{\mu}_b) + \Delta_8^{+}(\tilde{\mu}_b)]+ O(\tilde{\mu}_b^{-1}) = 0, 
\end{eqnarray}
where we have defined $\Delta_{r}^{\pm}(\tilde{\mu}_b) = \psi^{\tilde{\epsilon}_b}_{\delta_b,1}(\tilde{\mu}_b + r) \pm \psi^{\tilde{\epsilon}_b}_{\delta_b,1}(\tilde{\mu}_b - r)$ for the values $r=4,8$. Substituting our ansatz \eqref{apm} for $\psi^{\tilde{\epsilon}_b}_{\delta_b,1}(\tilde{\mu}_b)$ into the above equation and assuming that $\chi(\tilde{\mu}_b)$ admits a series expansion 
\begin{equation}
\chi(\tilde{\mu}_b \pm r)=\chi(\tilde{\mu}_b)\pm r\partial_{\tilde{\mu}_b}\chi|_{\tilde{\mu}_b}+\frac{r^2}{2}\partial^2_{\tilde{\mu}_b}\chi|_{\tilde{\mu}_b} + \cdots
\end{equation}
where each derivative with respect to ${\tilde{\mu}_b}$ decreases the asymptotic order in a unit, we finally obtain the following differential equation for any $\delta_b\neq 0$, up to terms of order $O(\tilde{\mu}_b^{-1})$:
\begin{equation}
\label{eq: App_de}
(1 + \delta_b^2 )[\tilde{\mu}_b^2\partial^2_{\tilde{\mu}_b}\chi (\tilde{\mu}_b) + \chi(\tilde{\mu}_b)] + 3(1 + \delta_b^2 )\tilde{\mu}_b\partial_{\tilde{\mu}_b}\chi (\tilde{\mu}_b) + m_c^2 \chi(\tilde{\mu}_b) = 0.
\end{equation} 
If we introduce in this expression the power law $\chi(\tilde{\mu}_b)=\tilde{\mu}_b^d$, we obtain for $d$ the two solutions given in Eq. \eqref{eq: IV_d}.

\end{document}